\documentclass[aps,pra,groupedaddress,twocolumn,showpacs]{revtex4-1}
\usepackage{amsmath}    
\usepackage{amsfonts}
\usepackage{bm}
\usepackage{amssymb}
\usepackage{appendix}
\usepackage{graphicx}   
\usepackage{color}      
\usepackage{subfigure}  
\usepackage{comment}
\usepackage{enumitem}

\newcommand{\re}{\textrm{Re}}
\newcommand{\im}{\textrm{Im}}

\begin{document}

\title{Eigenvalue dynamics in the presence of non-uniform gain and loss}

\author{Alexander Cerjan and Shanhui Fan} 
\affiliation{Department of Electrical Engineering, and Ginzton Laboratory, Stanford  University,  Stanford,  California  94305,  USA}


\date{\today}


\begin{abstract}
Loss-induced transmission in waveguides, and reversed pump dependence in lasers, 
are two prominent examples of counter-intuitive effects in non-Hermitian systems 
with patterned gain and loss. By analyzing the eigenvalue dynamics of complex symmetric 
matrices when a system parameter is varied, we introduce a general set of theoretical conditions
for these two effects. We show that these effects arise in any irreducible system where 
the gain or loss is added to a subset of the elements of the system, without the need 
for parity-time symmetry or for the system to be near
an exceptional point.
These results are confirmed using full-wave numerical simulations.
The conditions presented here vastly expand the design 
space for observing these effects. We also show that a similarly broad class of systems 
exhibit a loss-induced narrowing of the density of states. 
\end{abstract}


\maketitle

Recently, the study of parity-time ($\mathcal{PT}$) symmetric
optical systems has highlighted the importance of exploring non-Hermitian systems with patterned gain and loss \cite{bender_pt-symmetric_1999,bender_complex_2002,musslimani_optical_2008,makris_beam_2008,klaiman_visualization_2008,longhi_bloch_2009,makris_pt-symmetric_2010,ruter_observation_2010,chong_pt-symmetry_2011,ge_conservation_2012,hodaei_parity-time_symmetric_2014},
and has led to the discovery of a remarkable array of phenomena,
such as loss-induced transmission in waveguides \cite{guo_observation_2009},
unidirectional transport behavior \cite{lin_unidirectional_2011,regensburger_parity-time_2012,feng_experimental_2013,peng_parity-time-symmetric_2014,chang_parity-time_2014}, 
reversed pump dependence in lasers \cite{liertzer_pump-induced_2012,brandstetter_reversing_2014,peng_loss-induced_2014}, and 
band flattening in periodic structures \cite{makris_beam_2008,szameit_pt-symmetry_2011,ramezani_exceptional-point_2012,zhen_spawning_2015,cerjan_zipping_2016}.
These effects are leading to new possibilities for constructing on-chip 
integrated photonic circuits for the manipulation of light. 

Here, we focus on two of these effects: loss-induced transmission in waveguides, and reversed pump 
dependence in lasers. Both of these effects are fascinating since they are quite counter-intuitive.
They are also of potential practical importance in providing novel mechanisms for optical 
switching based on gain or loss modulation. In loss-induced transmission,
loss is added to one of two otherwise identical, parallel coupled waveguides \cite{guo_observation_2009}. After a critical amount of loss
is added, further
increases in the absorption also increases the total transmission through the waveguide pair.
Likewise, reversed pump dependence can be observed in laser systems consisting of two
coupled cavities \cite{liertzer_pump-induced_2012,brandstetter_reversing_2014,peng_loss-induced_2014,chitsazi_experimental_2014}. First, one of these cavities is pumped such that the total system begins
to lase. However, if the pump in the first cavity is then held constant and the gain
in the second cavity is increased, the total output lasing power 
can be seen to decrease
until the total gain distribution becomes relatively uniform,
so long as the added gain is sufficiently greater than the losses of the unpumped system.
If the laser is close to threshold after gain is added to the first cavity, 
this mechanism can drive the laser below threshold.

Initially, these types of counter-intuitive effects were found in $\mathcal{PT}$ symmetric
systems in their broken phase, and thus the onset of these behaviors was associated with the occurrence of an
exceptional point \cite{guo_observation_2009}. Subsequent analyses demonstrated that
loss-induced transmission in waveguides and reverse pump dependence in lasers could be found in systems which
were in the vicinity of an exceptional point, without requiring that the study system contain an exceptional point \cite{liertzer_pump-induced_2012,brandstetter_reversing_2014,peng_loss-induced_2014}.
A few recent studies have also interpreted reverse pump dependence in specific laser systems as the result of a loss or gain induced
impedance mis-matching, resulting in modes localized to specific regions of the system \cite{chitsazi_experimental_2014,longhi_wrong_arxiv},
and have argued that neither an exceptional point nor $\mathcal{PT}$ symmetry breaking are required in order to observe these behaviors.
However, given the significance of these counter-intuitive behaviors,
it is important to develop a set of general theoretical conditions for a system to exhibit these effects.



In this Rapid Communication, by analyzing the eigenvalue dynamics of the system matrix under modulation, 
we theoretically prove that both of these phenomena arise in a general class of structures 
in the presence of non-uniform gain and loss. 
In particular, loss-induced transmission can be observed in 
a class of linear systems with dimension no less than two, provided that the following
conditions are satisfied:
\begin{enumerate}[noitemsep,nolistsep,label=\arabic*)]
\item The underlying system matrix is irreducible when expressed on the basis set of individual waveguide modes.
\item The additional loss modulation is applied only to a subset of the waveguides.
\end{enumerate}
A very similar set of theoretical conditions can be developed for 
reversed pump dependence in lasers, which are then verified using full-wave simulations.
The conditions here are applicable to the known systems which display 
these effects, but moreover point to a much wider class of systems which exhibit
these behaviors without $\mathcal{PT}$ symmetry, 
or a requirement for the system to be close to an exceptional point.
Finally,
we show that systems undergoing a spatially non-uniform modulation of loss can also display a
loss-induced Purcell effect, where the density of states spectrum of the system narrows as 
the loss increases. This has potential applications in building narrow-band absorbing and emitting structures.

\begin{figure}[t!]
  \centering
  \includegraphics[width=0.48\textwidth]{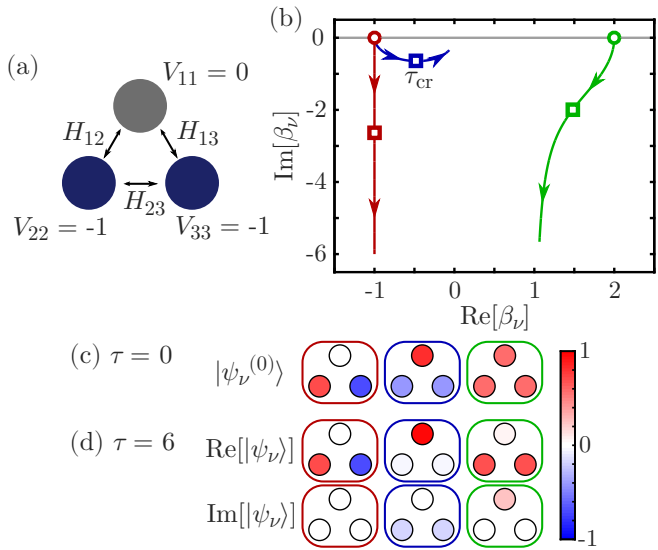}
  \caption{
    (a) Schematic of three equally coupled, $H_{nm}^{(0)} = 1$ for $n \ne m$, identical and otherwise lossless,
    $H_{nn}^{(0)}=0$, waveguides with $V_{11} = 0$, $V_{22} = V_{33} = -1$.
    (b) Motion of the eigenvalues, $\beta_\nu$,
    through the complex plane as $\tau \in [0,6]$ is increased, as indicated by the arrows. The circles (squares)
    indicate the locations of the eigenvalues when $\tau=0$ ($\tau = \tau_{\textrm{cr}} \approx 2.64$). 
    (c),(d) Representation of the modes of the three-waveguide systems when $\tau=0$
    (c), and $\tau=6$ (d). Eigenstate border colors
    correspond to the eigenvalue trajectories in (b). \label{fig:wg}}
\end{figure}

To illuminate the general theoretical principles, we will first consider an example of loss-induced 
transmission in a system consisting of three equally coupled, identical, single-mode waveguides 
in which two of the waveguides can have their loss tuned externally, shown in Fig.~\ref{fig:wg}(a).
The system dynamics can then be written in terms of the individual waveguide modes using coupled mode theory as \cite{haus}
\begin{equation}
\frac{1}{i}\frac{d}{dz} |c(z)\rangle = H|c(z)\rangle = \left(H^{(0)} + i\tau V\right) |c(z)\rangle. \label{eq:cmt1}
\end{equation}
Here, $c_n(z)$ is the modal amplitude on the $n$th waveguide, and the evolution
of the total system is given by $H$.
The underlying system is described by the symmetric matrix $H^{(0)}$, in which the diagonal entries represent
the propagation constants of the decoupled waveguides, and the off-diagonal entries
represent the coupling strengths between the waveguides. Furthermore, we assume
that $H^{(0)}$ is not block-diagonal, i.e.\ irreducible.
The additional loss modulation is described by $i\tau V$, where $V$ is a real diagonal matrix representing the
relative strength of the tunable loss in each waveguide, and $\tau \ge 0$ provides
an overall scaling to the loss. For the system shown in Fig.~\ref{fig:wg}(a), $V_{11} = 0$,
while $V_{22} = V_{33} = -1$. Thus, the additional loss is only applied to a 
sub-space in the system, spanned by two of the individual waveguide modes. 
This system therefore satisfies our stated conditions for loss-induced transmission.

The system in Fig.~\ref{fig:wg}(a) indeed exhibits loss-induced transmission. 
To illustrate, we set $|c(z)\rangle = e^{i \beta z}|\psi \rangle$, and solve Eq.~(\ref{eq:cmt1})
for the propagation eigenvalues, $\beta_\nu$, and corresponding eigenstates, $|\psi_\nu \rangle$, of
the total system as we vary the loss modulation strength $\tau$.
The eigenvalue trajectories for this system are shown in Fig.~\ref{fig:wg}(b) for $\tau \in [0,6]$. 
At $\tau = 0$, all of the eigenvalues reside on the real axis. Then, as $\tau$ is increased from zero, all 
of the eigenvalues initially move into the
negative half of the complex plane, demonstrating that all of the modes of the system
experience increasing loss for increasing $\tau$. However, beyond a critical value,
$\tau_{\textrm{cr}}$, one of the eigenvalues begins to return to the real axis. 
Thus, for values of $\tau > \tau_{\textrm{cr}}$, this system exhibits loss-induced transmission,
as the total transmission through the system becomes dominated by the transmission
through the lossless waveguide.

Previously, the onset of loss-induced transmission at $\tau = \tau_{\textrm{cr}}$
has been associated with the 
proximity of the system to, though not necessarily at,
an exceptional point,
where two of the eigenvalues of the system coincide and their corresponding 
eigenvectors become identical and self-orthogonal \cite{guo_observation_2009}.
Certainly, if one enlarged the parameter space by varying another
component(s) of the system, it is possible to find exceptional points.
For example, the system shown in Fig.~\ref{fig:wg}(a) can be tuned to contain
an exceptional point by decoupling the two lossy waveguides, $H_{23}^{(0)} = H_{32}^{(0)} = 0$,
or detuning the lossless waveguide from the lossy waveguides, $H_{11}^{(0)} = 1$.
The eigenvalue trajectories for these two systems as $\tau$ is increased are shown in the supplementary material \cite{supp_mat},
and other possible permutations containing exceptional points are likely to exist.
It is conceivable that the loss-induced transmission seen in Fig.~\ref{fig:wg}(b) can be
explained using an analytic continuation in parameter space from an exceptional point(s),
but such a theory has not yet been established. Furthermore, the analysis presented here
provides a simpler explanation for the observed loss-induced transmission.




Motivated by the specific example above, we now provide a proof of the general theoretical 
condition for systems exhibiting loss-induced transmission 
independent of any symmetry considerations.
We consider a general system as described by Eq.~(\ref{eq:cmt1})
with $N$ coupled elements. The underlying system is
assumed to either be lossless, or contain an overall uniform loss (or gain), such that
$H^{(0)} = \re[H^{(0)}] + i \delta I$, with $\re[H^{(0)}]$ symmetric. 
Thus, the eigenstates of $H^{(0)}$, $|\psi_\nu^{(0)}\rangle$, are entirely determined by $\re[H^{(0)}]$,
\begin{equation}
\re[H^{(0)}] |\psi_\nu^{(0)} \rangle = \re[\beta_\nu^{(0)}] |\psi_\nu^{(0)} \rangle,
\end{equation}
and its eigenvalues, $\beta_\nu^{(0)}$, all lie on the line $\im[\beta_\nu^{(0)}] = \delta$ in the complex plane. 
As loss is only added to a sub-space of the system, 
$V$ is negative semi-definite in addition to being real diagonal.
Moreover, by our condition 2 above, $V$ has a non-trivial kernel. Defining $K = \dim[\ker[V]]$, 
we then have $K \ge 1$. Here, $K$ represents the number of elements in the system that are 
individually unaffected by the addition of loss. 
To both avoid assumptions upon the magnitude of the elements of $H^{(0)}$
and for semantic convenience, we will use small and large values of $\tau$ to
refer to the regimes where either the term $H^{(0)}$ or the term $i\tau V$ dominates 
in $H$, respectively. 

To demonstrate that the system as described above always exhibits loss-induced transmission,
we will prove three propositions:
\begin{enumerate}[noitemsep,nolistsep,label=\arabic*)]
\item For small $\tau$, at least a subset, if not the full set,
of the eigenvalues of $H$, which satisfy $\im[\beta_\nu] = \delta$ at $\tau=0$, descend
from this line as $\tau$ is increased. Let $L$ be the number of elements of this subset, $L \le N$.
\item For large $\tau$, there are $K$ eigenvalues of $H$ which
remain on, or return to, the line $\im[\beta_\nu] = \delta$.
\item The number of eigenvalues which remain on the line $\im[\beta_\nu] = \delta$ as $\tau$ is initially increased
is strictly less than $K$, i.e.\ $N-L < K$.
\end{enumerate}
With these three propositions, at least one of the system's eigenvalues which is initially affected by the addition
of loss must eventually return to the line $\im[\beta_\nu] = \delta$, resulting in loss-induced transmission.

When $\tau$ is small, the eigenvalues of $H$ can be calculated perturbatively 
from the eigenvalues and eigenstates of $H^{(0)}$ as
\begin{equation}
\beta_\nu \approx \re[\beta_\nu^{(0)}] + i\delta + i \tau \langle \psi_\nu^{(0)} | V | \psi_\nu^{(0)} \rangle.
\end{equation}
As $V$ is negative semi-definite, there will be $L \le N$ eigenvalues for which
$|\psi_\nu^{(0)}\rangle \not\in \ker[V]$ such that $\im[\beta_\nu] < \delta$ as $\tau$
is initially increased from zero. Thus we have proved proposition 1 above.

When $\tau$ is large, the eigenvalues of the system can be calculated
perturbatively from the eigenvalues, $v_\nu$, and eigenstates, $|\varphi_\nu\rangle$ of $V$, as
\begin{equation}
\beta_\nu \approx i \tau v_\nu + \langle \varphi_\nu|\re[H^{(0)}]| \varphi_\nu \rangle + i\delta.
\end{equation}
As $V$ is both real diagonal and negative semi-definite, $v_\nu \le 0$, $\im[v_\nu] = 0$, and $|\varphi_\nu \rangle \in \mathbb{R}$. 
Thus, $\im[\beta_\nu] \to -\infty$ as $\tau \to \infty$ unless $v_\nu = 0$. 
But, as there are $K$ eigenvalues of $V$ with $v_\nu = 0$, there must be 
$K$ eigenvalues of $H$ with $\im[\beta_\nu] = \delta$.
Thus we have proved proposition 2 above.
The observed splitting of eigenvalues in the large $\tau$ limit
has been previously referred to as modal dichotomy \cite{figotin_dissipative_2012,figotin_lagrangian_2014}.
Modal dichotomy theory also predicts that transmitting eigenstates
exist in $\ker[V]$ for large $\tau$, and avoid the absorption in the system, while decaying
eigenstates cannot overlap with the lossless regions, confirmed here in Fig.\ \ref{fig:wg}(d).
This predicted behavior is similar to the eigenstates observed 
in $\mathcal{PT}$ symmetric systems in their broken phase \cite{ruter_observation_2010}.


At this point, we are guaranteed that $N-L \le K$, but
for loss-induced transmission to occur at large $\tau$, we must show
$N-L < K$.
Suppose $N-L = K$. Then there are $K$ eigenstates of $H^{(0)}$ which
exist in $\ker[V]$. Moreover, as the eigenstates of $H^{(0)}$ are linearly
independent, these $K$ eigenstates must span $\ker[V]$. This
means that $H^{(0)}$ can be written as a block diagonal matrix with at least
two blocks.
However, this contradicts our assumption that $H^{(0)}$ is irreducible. Therefore, 
there must always be
fewer than $K$ eigenstates of $H^{(0)}$ which exist in $\ker[V]$.
As such, at least one of the
eigenstates of $H$ which is initially not in the kernel of $V$ must evolve
into an eigenstate which is in the kernel of $V$ as $\tau$ increases, and exhibit loss-induced transmission.
This proof shows that the onset of loss-induced transmission
can be found in a general class of systems, and without
searching the parameter space of these systems for exceptional points.

This same argument can be used to show that reversed pump dependence \cite{liertzer_pump-induced_2012,brandstetter_reversing_2014,peng_loss-induced_2014} 
can be found in laser systems with at least two coupled cavities, provided
that they satisfy the conditions:
\begin{enumerate}[noitemsep,nolistsep,label=\arabic*)]
\item The underlying system is irreducible when expressed on a basis set of individual cavity modes.
\item The gain is added in two parts. First, gain is only increased in a subset of the coupled cavities. 
Second, gain is added to the remainder of the cavities in system, 
until the gain in the system is uniform.
\item Sufficient gain is added to reach the large $\tau$ limit.
\end{enumerate}
To demonstrate the connection with loss induced transmission in waveguides,
we decompose a laser system satisfying the above criteria into an underlying system with
large uniform gain as described by a matrix $H^{(0)}$ with $\delta > 0$, and an additional loss modulation term $i \tau V$. 
As reversed pump dependence can be equivalently described as the \textit{increase} of lasing power 
with the increase of the loss strength $\tau$ for some range of $\tau$, it will naturally occur provided 
that both $H^{(0)}$ and $V$ satisfy the conditions prescribed for loss induced transmission. To 
see that conditions 2 and 3 above are consistent with the second condition for loss induced transmission, 
we note that by condition 2 the loss modulation on top of the underlying lasing system is applied only 
to a subset of the cavities. Furthermore, condition 3 ensures that when only a subset
of the system has gain, i.e.\ after the `first' gain has been added but before the `second,' the system is in the large $\tau$ limit.
Therefore, any lasing system satisfying conditions 1-3 above must exhibit reversed pump dependence. 

\begin{figure}[t!]
  \centering
  \includegraphics[width=0.47\textwidth]{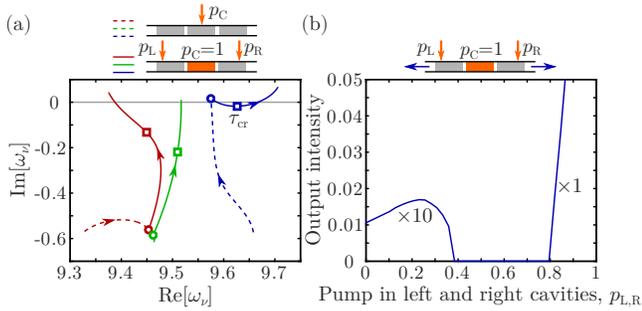}
  \caption{(a) Motion of the resonances, $\omega_\nu$, for three coupled cavities, each
    with $L = 1$mm and unpumped refractive index $n_{\textrm{bg}} = 3 + 0.163i$, separated by air gaps with $L = 100 \mu$m,
    as the total gain is increased. First, the gain in the center cavity is increased
    to $n_{\textrm{fp}} = 3 - 0.037i$ holding the gain in the left and right cavities constant, $p_{\textrm{C}} \in [0,1]$, $p_{\textrm{L,R}} = 0$ (dashed lines).
    Then, the gain in the center cavity is held constant while the left and right cavities
    are pumped to $n_{\textrm{fp}}$, $p_{\textrm{C}} = 1$, $p_{\textrm{L,R}} \in [0,1]$ (solid lines).
    The circles (squares) show the locations of the resonances at 
    $p_{\textrm{C}} = 1$, $p_{\textrm{L,R}} = 0$ ($p_{\textrm{C}} = 1$, $p_{\textrm{L,R}} = 0.6$, which corresponds to $\tau_{\textrm{cr}}$). 
    (b) SALT simulations of the same three cavity system showing the total output intensity 
    as the pump in the left and right cavities is increased from $p_{\textrm{L,R}} \in [0,1]$. The gain
    medium is assumed to be homogeneously broadened, with a central frequency of $k_{\textrm{a}} = 9.6$mm$^{-1}$, and width $\gamma_\perp = 0.2$mm$^{-1}$.
    The pump level is chosen to realize the refractive indexes from (a), 
    and $n_{\textrm{fp}}$ corresponds to $D_0 = 1.2$ in SALT units \cite{tureci06,ge10,cerjan_csalt_2015}.
    \label{fig:laser}}
\end{figure}

In applying the conditions above for real laser systems we have made two simplifications,
that the addition of uniform gain effects all of the system's modes equally, and that
the system is linear, which we must now address. The first simplification was made to ensure
that all of the resonances of the underlying system, $\omega_\nu^{(0)}$, lie along the line
$\im[\omega_\nu^{(0)}] = \delta$. However, in many cases this condition can be relaxed.
To demonstrate this,
we perform finite-difference frequency-domain (FDFD) simulations of a one-dimensional system
consisting of three cavities of equal lengths, coupled together by air gaps in between them,
as shown in Fig.~\ref{fig:laser}. First, only the center cavity is pumped, decreasing the imaginary
part of the refractive index from the background in the system, $n_{\textrm{bg}}$, to its fully pumped value, $n_{\textrm{fp}}$. This process is parameterized by 
$p_{\textrm{C}} = \im[n_{\textrm{C}}-n_{\textrm{bg}}]/\im[n_{\textrm{bg}}] \in [0,1]$, where $n_{\textrm{C}}$ is the refractive index
of the center cavity. Then, the left and right cavities are pumped, parameterized by $p_{\textrm{L,R}} \in [0,1]$
which are similarly defined, 
until the gain in all three cavities is uniform, $p_{\textrm{C}} = p_{\textrm{L,R}} = 1$. 
The motion of the resonances of the system through this pumping process is shown in Fig.~\ref{fig:laser}(a).
As can be seen, the system begins to lase, $\im[\omega_\nu] > 0$, when only the center cavity is pumped, $p_{\textrm{C}} = 1$ and $p_{\textrm{L,R}} = 0$, but falls
below threshold as the gain in the left and right cavities is increased. For example, all the eigenvalues have $\im[\omega_\nu]<0$ for $p_{\textrm{C}} = 1$ and $p_{\textrm{L,R}} = 0.6$. 
When additional gain is added to the left and right cavities, the system starts lasing again. 

Second, the underlying proof for reverse pump dependence assumes a linear system. 
However, lasers are an inherently non-linear system in which
gain saturation both prevents any resonance from moving into the positive half of the
complex plane, and also reduces the gain available for additional resonances to reach threshold when
the system is lasing. We can account for the effects of gain saturation within the FDFD
simulations by using the steady-state \textit{ab initio} laser theory (SALT),
which solves the set of self-consistent equations for each steady-state lasing mode coupled together through the
saturable gain medium \cite{tureci06,ge10,cerjan_csalt_2015}. SALT simulations of the three coupled cavity
system confirm that lasing ceases
and then resumes as gain is added to the left and right cavities, as shown in Fig.~\ref{fig:laser}(b).
However, the SALT simulations demonstrate that only a single lasing mode is above threshold 
when the gain in the cavity is uniform, $p_{\textrm{C}} = p_{\textrm{L,R}} = 1$, rather than all three modes as is
predicted in Fig.~\ref{fig:laser}(a). This is due to the gain saturation, which suppresses
the motion of the remaining resonances from reaching threshold for the pump values shown.

Together, the theoretical analysis and numerical simulations presented here verify that this system exhibits reverse pump dependence,
again without requiring a search of the system parameter space for exceptional points.
Thus, two simplifications made in order to arrive at the general theoretical conditions for reversed pump dependence 
amount to quantitative, not qualitative, corrections,
indicating that the simple but intuitive model presented here contains
the relevant physics for understanding both loss-induced transmission in waveguides and reverse pump dependence in lasers.
While it has been previously shown that exceptional point dynamics are not required
to exhibit reverse pump dependence in 
loss-coupled distributed-feedback
lasers \cite{longhi_wrong_arxiv}, the proof presented
here demonstrates the breadth of systems which can exhibit this behavior.

\begin{figure}[t!]
  \centering
  \includegraphics[width=0.48\textwidth]{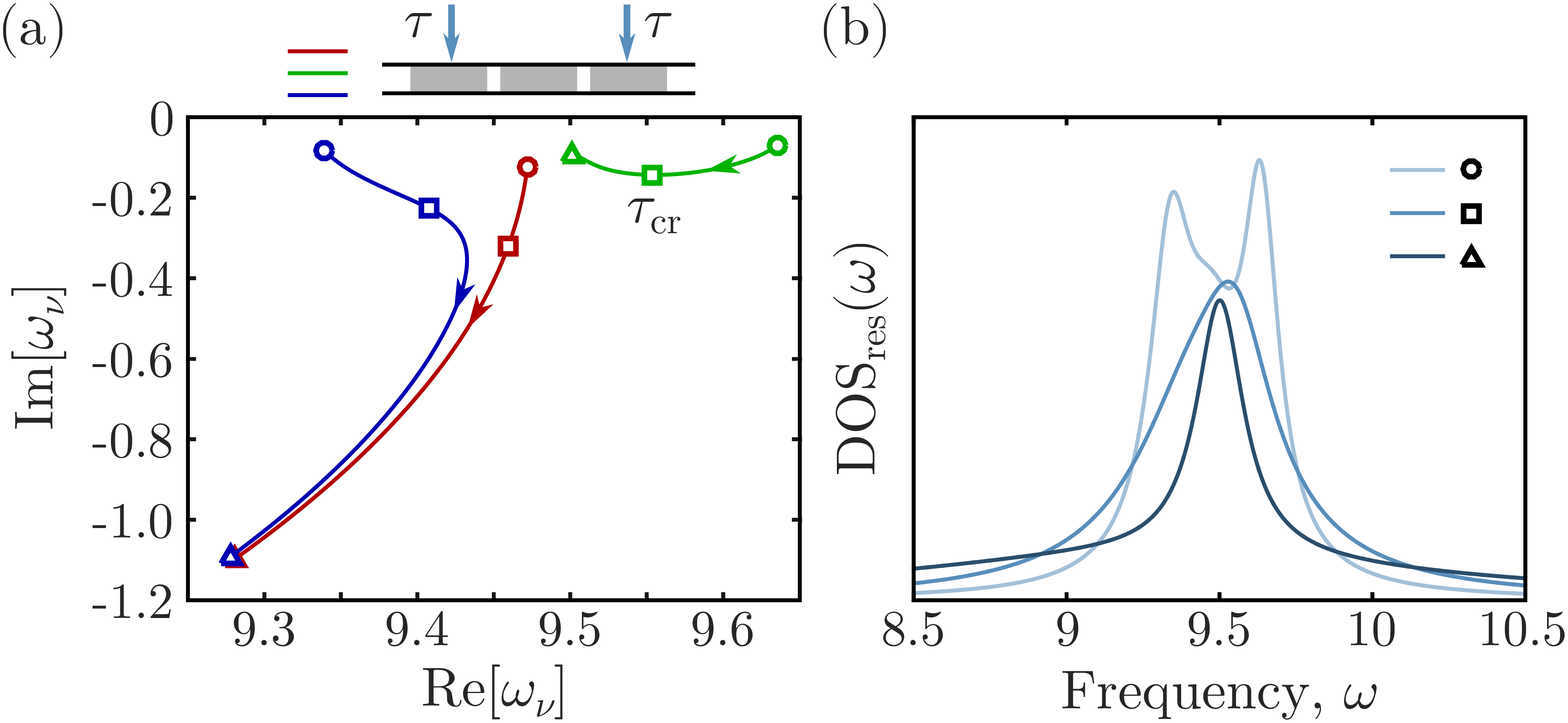}
  \caption{(a) Motion of the resonances, $\omega_\nu$, for three coupled cavities, each
    with $L = 1$mm and refractive index $n = 3 + 0.005i$, separated by air gaps with $L = 125 \mu$m,
    as the loss in the left and right cavities is increased to $n = 3.02+0.336i$. The circles show the locations
    of the resonances without any additional loss. The squares show the locations of the resonances
    at $\tau_{\textrm{cr}}$, which corresponds to a refractive index in the left and right cavities of $n = 3+0.072i$.
    The triangles show the locations of the resonances when the refractive index in the left and right cavities
    is $n = 3.02+0.336i$.
    (b) Plot of the contribution to the density of states, Eq.~(\ref{eq:dos}), from the three cavity resonances shown
    in part (a). The curves show the density of states with the locations of the resonances at the circles (light blue),
    squares (blue), and triangles (dark blue).
    \label{fig:dos}}
\end{figure}

The results above indicate that the new physical effects of loss-induced transmission and reversed pump 
dependence are in general related to spatially non-uniform gain/loss modulation in a photonic 
structure. We now show that such non-uniform gain/loss modulation may also have important ramifications 
on the density of states of a photonic structure. In general, the density of states of
a system is proportional to the imaginary part of the Green's function, which
is dominated by the resonances of the structure,
\begin{equation}
\textrm{DOS}_{\textrm{res}}(\omega) \sim \sum_{\nu} \frac{\gamma_\nu}{(\omega - \re[\omega_\nu])^2 + \gamma_\nu^2}, \label{eq:dos}
\end{equation}
where $\gamma_\nu = -\im[\omega_\nu]$ is the width of the resonance.
When a uniform loss is added to a system, all of the $\gamma_\nu$ increase,
resulting in the broadening of $\textrm{DOS}_{\textrm{res}}(\omega)$,
and $\textrm{DOS}_{\textrm{res}}(\omega) \to 0$ as $\gamma_\nu \to \infty$.
However, as we have demonstrated above, by adding a non-uniform loss modulation, $i \tau V$, 
to a loss-less underlying system, $H^{(0)}$,
at least one of the resonances which initially broadens for small $\tau$ must
return to the real axis for large $\tau$, such that $\textrm{DOS}_{\textrm{res}}(\omega) > 0$
even as $\tau \to \infty$, thus exhibiting loss-induced narrowing of the
density of states spectrum for some range of $\tau$.
Another way to view this is as a loss-induced Purcell effect due to the localization of the
eigenstates.

This narrowing of the density of states can be seen in a one-dimensional system consisting
of three coupled cavities surrounded by air, depicted in Fig.\ \ref{fig:dos}(a). As the loss is increased
in the left and right cavities, two of the resonances of the system are seen to descend deep
into the lower half of the complex plane, while the remaining resonance initially descends, but then begins
to ascend for $\tau > \tau_{\textrm{cr}}$. During this process, the density of states spectrum narrows 
as the contributions from the two diverging resonances diminish, and as the remaining resonance returns
towards the real axis, as shown in Fig.\ \ref{fig:dos}(b).

In conclusion, we have provided the general theoretical conditions for finding
loss-induced transparency in waveguides and reverse pump dependence in lasers.
We show that these effects generally arise in a class of systems with non-uniformly modulated gain/loss,
and that 
these systems need not have parity-time symmetry or be close to exceptional points. Furthermore, the same class of systems 
can also exhibit a narrowing of the density of state spectrum with increasing loss. 
Our results demonstrate the breadth of potential device designs with
non-uniform gain and loss which can be used to create counter-intuitive optical effects
with potentially important applications.


\begin{acknowledgments}
We would like to thank Steven Johnson, Adi Pick, Matthias Liertzer, and Sacha Vers for stimulating
discussions.
This work was supported by an AFOSR MURI program (Grant
No.\ FA9550-12-1-0471), and an AFOSR project (Grant No.\ FA9550-16-1-0010). 
\end{acknowledgments}


\begin{thebibliography}{33}%
\makeatletter
\providecommand \@ifxundefined [1]{%
 \@ifx{#1\undefined}
}%
\providecommand \@ifnum [1]{%
 \ifnum #1\expandafter \@firstoftwo
 \else \expandafter \@secondoftwo
 \fi
}%
\providecommand \@ifx [1]{%
 \ifx #1\expandafter \@firstoftwo
 \else \expandafter \@secondoftwo
 \fi
}%
\providecommand \natexlab [1]{#1}%
\providecommand \enquote  [1]{``#1''}%
\providecommand \bibnamefont  [1]{#1}%
\providecommand \bibfnamefont [1]{#1}%
\providecommand \citenamefont [1]{#1}%
\providecommand \href@noop [0]{\@secondoftwo}%
\providecommand \href [0]{\begingroup \@sanitize@url \@href}%
\providecommand \@href[1]{\@@startlink{#1}\@@href}%
\providecommand \@@href[1]{\endgroup#1\@@endlink}%
\providecommand \@sanitize@url [0]{\catcode `\\12\catcode `\$12\catcode
  `\&12\catcode `\#12\catcode `\^12\catcode `\_12\catcode `\%12\relax}%
\providecommand \@@startlink[1]{}%
\providecommand \@@endlink[0]{}%
\providecommand \url  [0]{\begingroup\@sanitize@url \@url }%
\providecommand \@url [1]{\endgroup\@href {#1}{\urlprefix }}%
\providecommand \urlprefix  [0]{URL }%
\providecommand \Eprint [0]{\href }%
\providecommand \doibase [0]{http://dx.doi.org/}%
\providecommand \selectlanguage [0]{\@gobble}%
\providecommand \bibinfo  [0]{\@secondoftwo}%
\providecommand \bibfield  [0]{\@secondoftwo}%
\providecommand \translation [1]{[#1]}%
\providecommand \BibitemOpen [0]{}%
\providecommand \bibitemStop [0]{}%
\providecommand \bibitemNoStop [0]{.\EOS\space}%
\providecommand \EOS [0]{\spacefactor3000\relax}%
\providecommand \BibitemShut  [1]{\csname bibitem#1\endcsname}%
\let\auto@bib@innerbib\@empty
\bibitem [{\citenamefont {Bender}\ \emph {et~al.}(1999)\citenamefont {Bender},
  \citenamefont {Boettcher},\ and\ \citenamefont
  {Meisinger}}]{bender_pt-symmetric_1999}%
  \BibitemOpen
  \bibfield  {author} {\bibinfo {author} {\bibfnamefont {C.~M.}\ \bibnamefont
  {Bender}}, \bibinfo {author} {\bibfnamefont {S.}~\bibnamefont {Boettcher}}, \
  and\ \bibinfo {author} {\bibfnamefont {P.~N.}\ \bibnamefont {Meisinger}},\
  }\href {\doibase 10.1063/1.532860} {\bibfield  {journal} {\bibinfo  {journal}
  {J. Math. Phys.}\ }\textbf {\bibinfo {volume} {40}},\ \bibinfo {pages} {2201}
  (\bibinfo {year} {1999})}\BibitemShut {NoStop}%
\bibitem [{\citenamefont {Bender}\ \emph {et~al.}(2002)\citenamefont {Bender},
  \citenamefont {Brody},\ and\ \citenamefont {Jones}}]{bender_complex_2002}%
  \BibitemOpen
  \bibfield  {author} {\bibinfo {author} {\bibfnamefont {C.~M.}\ \bibnamefont
  {Bender}}, \bibinfo {author} {\bibfnamefont {D.~C.}\ \bibnamefont {Brody}}, \
  and\ \bibinfo {author} {\bibfnamefont {H.~F.}\ \bibnamefont {Jones}},\ }\href
  {\doibase 10.1103/PhysRevLett.89.270401} {\bibfield  {journal} {\bibinfo
  {journal} {Phys. Rev. Lett.}\ }\textbf {\bibinfo {volume} {89}},\ \bibinfo
  {pages} {270401} (\bibinfo {year} {2002})}\BibitemShut {NoStop}%
\bibitem [{\citenamefont {Musslimani}\ \emph {et~al.}(2008)\citenamefont
  {Musslimani}, \citenamefont {Makris}, \citenamefont {El-Ganainy},\ and\
  \citenamefont {Christodoulides}}]{musslimani_optical_2008}%
  \BibitemOpen
  \bibfield  {author} {\bibinfo {author} {\bibfnamefont {Z.~H.}\ \bibnamefont
  {Musslimani}}, \bibinfo {author} {\bibfnamefont {K.~G.}\ \bibnamefont
  {Makris}}, \bibinfo {author} {\bibfnamefont {R.}~\bibnamefont {El-Ganainy}},
  \ and\ \bibinfo {author} {\bibfnamefont {D.~N.}\ \bibnamefont
  {Christodoulides}},\ }\href {\doibase 10.1103/PhysRevLett.100.030402}
  {\bibfield  {journal} {\bibinfo  {journal} {Phys. Rev. Lett.}\ }\textbf
  {\bibinfo {volume} {100}},\ \bibinfo {pages} {030402} (\bibinfo {year}
  {2008})}\BibitemShut {NoStop}%
\bibitem [{\citenamefont {Makris}\ \emph {et~al.}(2008)\citenamefont {Makris},
  \citenamefont {El-Ganainy}, \citenamefont {Christodoulides},\ and\
  \citenamefont {Musslimani}}]{makris_beam_2008}%
  \BibitemOpen
  \bibfield  {author} {\bibinfo {author} {\bibfnamefont {K.~G.}\ \bibnamefont
  {Makris}}, \bibinfo {author} {\bibfnamefont {R.}~\bibnamefont {El-Ganainy}},
  \bibinfo {author} {\bibfnamefont {D.~N.}\ \bibnamefont {Christodoulides}}, \
  and\ \bibinfo {author} {\bibfnamefont {Z.~H.}\ \bibnamefont {Musslimani}},\
  }\href {\doibase 10.1103/PhysRevLett.100.103904} {\bibfield  {journal}
  {\bibinfo  {journal} {Phys. Rev. Lett.}\ }\textbf {\bibinfo {volume} {100}},\
  \bibinfo {pages} {103904} (\bibinfo {year} {2008})}\BibitemShut {NoStop}%
\bibitem [{\citenamefont {Klaiman}\ \emph {et~al.}(2008)\citenamefont
  {Klaiman}, \citenamefont {G\"{u}nther},\ and\ \citenamefont
  {Moiseyev}}]{klaiman_visualization_2008}%
  \BibitemOpen
  \bibfield  {author} {\bibinfo {author} {\bibfnamefont {S.}~\bibnamefont
  {Klaiman}}, \bibinfo {author} {\bibfnamefont {U.}~\bibnamefont
  {G\"{u}nther}}, \ and\ \bibinfo {author} {\bibfnamefont {N.}~\bibnamefont
  {Moiseyev}},\ }\href {\doibase 10.1103/PhysRevLett.101.080402} {\bibfield
  {journal} {\bibinfo  {journal} {Phys. Rev. Lett.}\ }\textbf {\bibinfo
  {volume} {101}},\ \bibinfo {pages} {080402} (\bibinfo {year}
  {2008})}\BibitemShut {NoStop}%
\bibitem [{\citenamefont {Longhi}(2009)}]{longhi_bloch_2009}%
  \BibitemOpen
  \bibfield  {author} {\bibinfo {author} {\bibfnamefont {S.}~\bibnamefont
  {Longhi}},\ }\href {\doibase 10.1103/PhysRevLett.103.123601} {\bibfield
  {journal} {\bibinfo  {journal} {Phys. Rev. Lett.}\ }\textbf {\bibinfo
  {volume} {103}},\ \bibinfo {pages} {123601} (\bibinfo {year}
  {2009})}\BibitemShut {NoStop}%
\bibitem [{\citenamefont {Makris}\ \emph {et~al.}(2010)\citenamefont {Makris},
  \citenamefont {El-Ganainy}, \citenamefont {Christodoulides},\ and\
  \citenamefont {Musslimani}}]{makris_pt-symmetric_2010}%
  \BibitemOpen
  \bibfield  {author} {\bibinfo {author} {\bibfnamefont {K.~G.}\ \bibnamefont
  {Makris}}, \bibinfo {author} {\bibfnamefont {R.}~\bibnamefont {El-Ganainy}},
  \bibinfo {author} {\bibfnamefont {D.~N.}\ \bibnamefont {Christodoulides}}, \
  and\ \bibinfo {author} {\bibfnamefont {Z.~H.}\ \bibnamefont {Musslimani}},\
  }\href {\doibase 10.1103/PhysRevA.81.063807} {\bibfield  {journal} {\bibinfo
  {journal} {Phys. Rev. A}\ }\textbf {\bibinfo {volume} {81}},\ \bibinfo
  {pages} {063807} (\bibinfo {year} {2010})}\BibitemShut {NoStop}%
\bibitem [{\citenamefont {R\"{u}ter}\ \emph {et~al.}(2010)\citenamefont
  {R\"{u}ter}, \citenamefont {Makris}, \citenamefont {El-Ganainy},
  \citenamefont {Christodoulides}, \citenamefont {Segev},\ and\ \citenamefont
  {Kip}}]{ruter_observation_2010}%
  \BibitemOpen
  \bibfield  {author} {\bibinfo {author} {\bibfnamefont {C.~E.}\ \bibnamefont
  {R\"{u}ter}}, \bibinfo {author} {\bibfnamefont {K.~G.}\ \bibnamefont
  {Makris}}, \bibinfo {author} {\bibfnamefont {R.}~\bibnamefont {El-Ganainy}},
  \bibinfo {author} {\bibfnamefont {D.~N.}\ \bibnamefont {Christodoulides}},
  \bibinfo {author} {\bibfnamefont {M.}~\bibnamefont {Segev}}, \ and\ \bibinfo
  {author} {\bibfnamefont {D.}~\bibnamefont {Kip}},\ }\href {\doibase
  10.1038/nphys1515} {\bibfield  {journal} {\bibinfo  {journal} {Nat. Phys.}\
  }\textbf {\bibinfo {volume} {6}},\ \bibinfo {pages} {192} (\bibinfo {year}
  {2010})}\BibitemShut {NoStop}%
\bibitem [{\citenamefont {Chong}\ \emph {et~al.}(2011)\citenamefont {Chong},
  \citenamefont {Ge},\ and\ \citenamefont {Stone}}]{chong_pt-symmetry_2011}%
  \BibitemOpen
  \bibfield  {author} {\bibinfo {author} {\bibfnamefont {Y.~D.}\ \bibnamefont
  {Chong}}, \bibinfo {author} {\bibfnamefont {L.}~\bibnamefont {Ge}}, \ and\
  \bibinfo {author} {\bibfnamefont {A.~D.}\ \bibnamefont {Stone}},\ }\href
  {\doibase 10.1103/PhysRevLett.106.093902} {\bibfield  {journal} {\bibinfo
  {journal} {Phys. Rev. Lett.}\ }\textbf {\bibinfo {volume} {106}},\ \bibinfo
  {pages} {093902} (\bibinfo {year} {2011})}\BibitemShut {NoStop}%
\bibitem [{\citenamefont {Ge}\ \emph {et~al.}(2012)\citenamefont {Ge},
  \citenamefont {Chong},\ and\ \citenamefont {Stone}}]{ge_conservation_2012}%
  \BibitemOpen
  \bibfield  {author} {\bibinfo {author} {\bibfnamefont {L.}~\bibnamefont
  {Ge}}, \bibinfo {author} {\bibfnamefont {Y.~D.}\ \bibnamefont {Chong}}, \
  and\ \bibinfo {author} {\bibfnamefont {A.~D.}\ \bibnamefont {Stone}},\ }\href
  {\doibase 10.1103/PhysRevA.85.023802} {\bibfield  {journal} {\bibinfo
  {journal} {Phys. Rev. A}\ }\textbf {\bibinfo {volume} {85}},\ \bibinfo
  {pages} {023802} (\bibinfo {year} {2012})}\BibitemShut {NoStop}%
\bibitem [{\citenamefont {Hodaei}\ \emph {et~al.}(2014)\citenamefont {Hodaei},
  \citenamefont {Miri}, \citenamefont {Heinrich}, \citenamefont
  {Christodoulides},\ and\ \citenamefont
  {Khajavikhan}}]{hodaei_parity-time_symmetric_2014}%
  \BibitemOpen
  \bibfield  {author} {\bibinfo {author} {\bibfnamefont {H.}~\bibnamefont
  {Hodaei}}, \bibinfo {author} {\bibfnamefont {M.-A.}\ \bibnamefont {Miri}},
  \bibinfo {author} {\bibfnamefont {M.}~\bibnamefont {Heinrich}}, \bibinfo
  {author} {\bibfnamefont {D.~N.}\ \bibnamefont {Christodoulides}}, \ and\
  \bibinfo {author} {\bibfnamefont {M.}~\bibnamefont {Khajavikhan}},\ }\href
  {\doibase 10.1126/science.1258480} {\bibfield  {journal} {\bibinfo  {journal}
  {Science}\ }\textbf {\bibinfo {volume} {346}},\ \bibinfo {pages} {975}
  (\bibinfo {year} {2014})}\BibitemShut {NoStop}%
\bibitem [{\citenamefont {Guo}\ \emph {et~al.}(2009)\citenamefont {Guo},
  \citenamefont {Salamo}, \citenamefont {Duchesne}, \citenamefont {Morandotti},
  \citenamefont {Volatier-Ravat}, \citenamefont {Aimez}, \citenamefont
  {Siviloglou},\ and\ \citenamefont {Christodoulides}}]{guo_observation_2009}%
  \BibitemOpen
  \bibfield  {author} {\bibinfo {author} {\bibfnamefont {A.}~\bibnamefont
  {Guo}}, \bibinfo {author} {\bibfnamefont {G.~J.}\ \bibnamefont {Salamo}},
  \bibinfo {author} {\bibfnamefont {D.}~\bibnamefont {Duchesne}}, \bibinfo
  {author} {\bibfnamefont {R.}~\bibnamefont {Morandotti}}, \bibinfo {author}
  {\bibfnamefont {M.}~\bibnamefont {Volatier-Ravat}}, \bibinfo {author}
  {\bibfnamefont {V.}~\bibnamefont {Aimez}}, \bibinfo {author} {\bibfnamefont
  {G.~A.}\ \bibnamefont {Siviloglou}}, \ and\ \bibinfo {author} {\bibfnamefont
  {D.~N.}\ \bibnamefont {Christodoulides}},\ }\href {\doibase
  10.1103/PhysRevLett.103.093902} {\bibfield  {journal} {\bibinfo  {journal}
  {Phys. Rev. Lett.}\ }\textbf {\bibinfo {volume} {103}},\ \bibinfo {pages}
  {093902} (\bibinfo {year} {2009})}\BibitemShut {NoStop}%
\bibitem [{\citenamefont {Lin}\ \emph {et~al.}(2011)\citenamefont {Lin},
  \citenamefont {Ramezani}, \citenamefont {Eichelkraut}, \citenamefont
  {Kottos}, \citenamefont {Cao},\ and\ \citenamefont
  {Christodoulides}}]{lin_unidirectional_2011}%
  \BibitemOpen
  \bibfield  {author} {\bibinfo {author} {\bibfnamefont {Z.}~\bibnamefont
  {Lin}}, \bibinfo {author} {\bibfnamefont {H.}~\bibnamefont {Ramezani}},
  \bibinfo {author} {\bibfnamefont {T.}~\bibnamefont {Eichelkraut}}, \bibinfo
  {author} {\bibfnamefont {T.}~\bibnamefont {Kottos}}, \bibinfo {author}
  {\bibfnamefont {H.}~\bibnamefont {Cao}}, \ and\ \bibinfo {author}
  {\bibfnamefont {D.~N.}\ \bibnamefont {Christodoulides}},\ }\href {\doibase
  10.1103/PhysRevLett.106.213901} {\bibfield  {journal} {\bibinfo  {journal}
  {Phys. Rev. Lett.}\ }\textbf {\bibinfo {volume} {106}},\ \bibinfo {pages}
  {213901} (\bibinfo {year} {2011})}\BibitemShut {NoStop}%
\bibitem [{\citenamefont {Regensburger}\ \emph {et~al.}(2012)\citenamefont
  {Regensburger}, \citenamefont {Bersch}, \citenamefont {Miri}, \citenamefont
  {Onishchukov}, \citenamefont {Christodoulides},\ and\ \citenamefont
  {Peschel}}]{regensburger_parity-time_2012}%
  \BibitemOpen
  \bibfield  {author} {\bibinfo {author} {\bibfnamefont {A.}~\bibnamefont
  {Regensburger}}, \bibinfo {author} {\bibfnamefont {C.}~\bibnamefont
  {Bersch}}, \bibinfo {author} {\bibfnamefont {M.-A.}\ \bibnamefont {Miri}},
  \bibinfo {author} {\bibfnamefont {G.}~\bibnamefont {Onishchukov}}, \bibinfo
  {author} {\bibfnamefont {D.~N.}\ \bibnamefont {Christodoulides}}, \ and\
  \bibinfo {author} {\bibfnamefont {U.}~\bibnamefont {Peschel}},\ }\href
  {\doibase 10.1038/nature11298} {\bibfield  {journal} {\bibinfo  {journal}
  {Nature}\ }\textbf {\bibinfo {volume} {488}},\ \bibinfo {pages} {167}
  (\bibinfo {year} {2012})}\BibitemShut {NoStop}%
\bibitem [{\citenamefont {Feng}\ \emph {et~al.}(2013)\citenamefont {Feng},
  \citenamefont {Xu}, \citenamefont {Fegadolli}, \citenamefont {Lu},
  \citenamefont {Oliveira}, \citenamefont {Almeida}, \citenamefont {Chen},\
  and\ \citenamefont {Scherer}}]{feng_experimental_2013}%
  \BibitemOpen
  \bibfield  {author} {\bibinfo {author} {\bibfnamefont {L.}~\bibnamefont
  {Feng}}, \bibinfo {author} {\bibfnamefont {Y.-L.}\ \bibnamefont {Xu}},
  \bibinfo {author} {\bibfnamefont {W.~S.}\ \bibnamefont {Fegadolli}}, \bibinfo
  {author} {\bibfnamefont {M.-H.}\ \bibnamefont {Lu}}, \bibinfo {author}
  {\bibfnamefont {J.~E.~B.}\ \bibnamefont {Oliveira}}, \bibinfo {author}
  {\bibfnamefont {V.~R.}\ \bibnamefont {Almeida}}, \bibinfo {author}
  {\bibfnamefont {Y.-F.}\ \bibnamefont {Chen}}, \ and\ \bibinfo {author}
  {\bibfnamefont {A.}~\bibnamefont {Scherer}},\ }\href {\doibase
  10.1038/nmat3495} {\bibfield  {journal} {\bibinfo  {journal} {Nat. Mater.}\
  }\textbf {\bibinfo {volume} {12}},\ \bibinfo {pages} {108} (\bibinfo {year}
  {2013})}\BibitemShut {NoStop}%
\bibitem [{\citenamefont {Peng}\ \emph
  {et~al.}(2014{\natexlab{a}})\citenamefont {Peng}, \citenamefont
  {\"{O}zdemir}, \citenamefont {Lei}, \citenamefont {Monifi}, \citenamefont
  {Gianfreda}, \citenamefont {Long}, \citenamefont {Fan}, \citenamefont {Nori},
  \citenamefont {Bender},\ and\ \citenamefont
  {Yang}}]{peng_parity-time-symmetric_2014}%
  \BibitemOpen
  \bibfield  {author} {\bibinfo {author} {\bibfnamefont {B.}~\bibnamefont
  {Peng}}, \bibinfo {author} {\bibfnamefont {{\c S}.~K.}\ \bibnamefont
  {\"{O}zdemir}}, \bibinfo {author} {\bibfnamefont {F.}~\bibnamefont {Lei}},
  \bibinfo {author} {\bibfnamefont {F.}~\bibnamefont {Monifi}}, \bibinfo
  {author} {\bibfnamefont {M.}~\bibnamefont {Gianfreda}}, \bibinfo {author}
  {\bibfnamefont {G.~L.}\ \bibnamefont {Long}}, \bibinfo {author}
  {\bibfnamefont {S.}~\bibnamefont {Fan}}, \bibinfo {author} {\bibfnamefont
  {F.}~\bibnamefont {Nori}}, \bibinfo {author} {\bibfnamefont {C.~M.}\
  \bibnamefont {Bender}}, \ and\ \bibinfo {author} {\bibfnamefont
  {L.}~\bibnamefont {Yang}},\ }\href {\doibase 10.1038/nphys2927} {\bibfield
  {journal} {\bibinfo  {journal} {Nat. Phys.}\ }\textbf {\bibinfo {volume}
  {10}},\ \bibinfo {pages} {394} (\bibinfo {year}
  {2014}{\natexlab{a}})}\BibitemShut {NoStop}%
\bibitem [{\citenamefont {Chang}\ \emph {et~al.}(2014)\citenamefont {Chang},
  \citenamefont {Jiang}, \citenamefont {Hua}, \citenamefont {Yang},
  \citenamefont {Wen}, \citenamefont {Jiang}, \citenamefont {Li}, \citenamefont
  {Wang},\ and\ \citenamefont {Xiao}}]{chang_parity-time_2014}%
  \BibitemOpen
  \bibfield  {author} {\bibinfo {author} {\bibfnamefont {L.}~\bibnamefont
  {Chang}}, \bibinfo {author} {\bibfnamefont {X.}~\bibnamefont {Jiang}},
  \bibinfo {author} {\bibfnamefont {S.}~\bibnamefont {Hua}}, \bibinfo {author}
  {\bibfnamefont {C.}~\bibnamefont {Yang}}, \bibinfo {author} {\bibfnamefont
  {J.}~\bibnamefont {Wen}}, \bibinfo {author} {\bibfnamefont {L.}~\bibnamefont
  {Jiang}}, \bibinfo {author} {\bibfnamefont {G.}~\bibnamefont {Li}}, \bibinfo
  {author} {\bibfnamefont {G.}~\bibnamefont {Wang}}, \ and\ \bibinfo {author}
  {\bibfnamefont {M.}~\bibnamefont {Xiao}},\ }\href {\doibase
  10.1038/nphoton.2014.133} {\bibfield  {journal} {\bibinfo  {journal} {Nat.
  Photonics}\ }\textbf {\bibinfo {volume} {8}},\ \bibinfo {pages} {524}
  (\bibinfo {year} {2014})}\BibitemShut {NoStop}%
\bibitem [{\citenamefont {Liertzer}\ \emph {et~al.}(2012)\citenamefont
  {Liertzer}, \citenamefont {Ge}, \citenamefont {Cerjan}, \citenamefont
  {Stone}, \citenamefont {T\"{u}reci},\ and\ \citenamefont
  {Rotter}}]{liertzer_pump-induced_2012}%
  \BibitemOpen
  \bibfield  {author} {\bibinfo {author} {\bibfnamefont {M.}~\bibnamefont
  {Liertzer}}, \bibinfo {author} {\bibfnamefont {L.}~\bibnamefont {Ge}},
  \bibinfo {author} {\bibfnamefont {A.}~\bibnamefont {Cerjan}}, \bibinfo
  {author} {\bibfnamefont {A.~D.}\ \bibnamefont {Stone}}, \bibinfo {author}
  {\bibfnamefont {H.~E.}\ \bibnamefont {T\"{u}reci}}, \ and\ \bibinfo {author}
  {\bibfnamefont {S.}~\bibnamefont {Rotter}},\ }\href {\doibase
  10.1103/PhysRevLett.108.173901} {\bibfield  {journal} {\bibinfo  {journal}
  {Phys. Rev. Lett.}\ }\textbf {\bibinfo {volume} {108}},\ \bibinfo {pages}
  {173901} (\bibinfo {year} {2012})}\BibitemShut {NoStop}%
\bibitem [{\citenamefont {Brandstetter}\ \emph {et~al.}(2014)\citenamefont
  {Brandstetter}, \citenamefont {Liertzer}, \citenamefont {Deutsch},
  \citenamefont {Klang}, \citenamefont {Sch\"{o}berl}, \citenamefont
  {T\"{u}reci}, \citenamefont {Strasser}, \citenamefont {Unterrainer},\ and\
  \citenamefont {Rotter}}]{brandstetter_reversing_2014}%
  \BibitemOpen
  \bibfield  {author} {\bibinfo {author} {\bibfnamefont {M.}~\bibnamefont
  {Brandstetter}}, \bibinfo {author} {\bibfnamefont {M.}~\bibnamefont
  {Liertzer}}, \bibinfo {author} {\bibfnamefont {C.}~\bibnamefont {Deutsch}},
  \bibinfo {author} {\bibfnamefont {P.}~\bibnamefont {Klang}}, \bibinfo
  {author} {\bibfnamefont {J.}~\bibnamefont {Sch\"{o}berl}}, \bibinfo {author}
  {\bibfnamefont {H.~E.}\ \bibnamefont {T\"{u}reci}}, \bibinfo {author}
  {\bibfnamefont {G.}~\bibnamefont {Strasser}}, \bibinfo {author}
  {\bibfnamefont {K.}~\bibnamefont {Unterrainer}}, \ and\ \bibinfo {author}
  {\bibfnamefont {S.}~\bibnamefont {Rotter}},\ }\href {\doibase
  10.1038/ncomms5034} {\bibfield  {journal} {\bibinfo  {journal} {Nat.
  Commun.}\ }\textbf {\bibinfo {volume} {5}},\ \bibinfo {pages} {4034}
  (\bibinfo {year} {2014})}\BibitemShut {NoStop}%
\bibitem [{\citenamefont {Peng}\ \emph
  {et~al.}(2014{\natexlab{b}})\citenamefont {Peng}, \citenamefont
  {\"{O}zdemir}, \citenamefont {Rotter}, \citenamefont {Yilmaz}, \citenamefont
  {Liertzer}, \citenamefont {Monifi}, \citenamefont {Bender}, \citenamefont
  {Nori},\ and\ \citenamefont {Yang}}]{peng_loss-induced_2014}%
  \BibitemOpen
  \bibfield  {author} {\bibinfo {author} {\bibfnamefont {B.}~\bibnamefont
  {Peng}}, \bibinfo {author} {\bibfnamefont {{\c S}.~K.}\ \bibnamefont
  {\"{O}zdemir}}, \bibinfo {author} {\bibfnamefont {S.}~\bibnamefont {Rotter}},
  \bibinfo {author} {\bibfnamefont {H.}~\bibnamefont {Yilmaz}}, \bibinfo
  {author} {\bibfnamefont {M.}~\bibnamefont {Liertzer}}, \bibinfo {author}
  {\bibfnamefont {F.}~\bibnamefont {Monifi}}, \bibinfo {author} {\bibfnamefont
  {C.~M.}\ \bibnamefont {Bender}}, \bibinfo {author} {\bibfnamefont
  {F.}~\bibnamefont {Nori}}, \ and\ \bibinfo {author} {\bibfnamefont
  {L.}~\bibnamefont {Yang}},\ }\href {\doibase 10.1126/science.1258004}
  {\bibfield  {journal} {\bibinfo  {journal} {Science}\ }\textbf {\bibinfo
  {volume} {346}},\ \bibinfo {pages} {328} (\bibinfo {year}
  {2014}{\natexlab{b}})}\BibitemShut {NoStop}%
\bibitem [{\citenamefont {Szameit}\ \emph {et~al.}(2011)\citenamefont
  {Szameit}, \citenamefont {Rechtsman}, \citenamefont {Bahat-Treidel},\ and\
  \citenamefont {Segev}}]{szameit_pt-symmetry_2011}%
  \BibitemOpen
  \bibfield  {author} {\bibinfo {author} {\bibfnamefont {A.}~\bibnamefont
  {Szameit}}, \bibinfo {author} {\bibfnamefont {M.~C.}\ \bibnamefont
  {Rechtsman}}, \bibinfo {author} {\bibfnamefont {O.}~\bibnamefont
  {Bahat-Treidel}}, \ and\ \bibinfo {author} {\bibfnamefont {M.}~\bibnamefont
  {Segev}},\ }\href {\doibase 10.1103/PhysRevA.84.021806} {\bibfield  {journal}
  {\bibinfo  {journal} {Phys. Rev. A}\ }\textbf {\bibinfo {volume} {84}},\
  \bibinfo {pages} {021806} (\bibinfo {year} {2011})}\BibitemShut {NoStop}%
\bibitem [{\citenamefont {Ramezani}\ \emph {et~al.}(2012)\citenamefont
  {Ramezani}, \citenamefont {Kottos}, \citenamefont {Kovanis},\ and\
  \citenamefont {Christodoulides}}]{ramezani_exceptional-point_2012}%
  \BibitemOpen
  \bibfield  {author} {\bibinfo {author} {\bibfnamefont {H.}~\bibnamefont
  {Ramezani}}, \bibinfo {author} {\bibfnamefont {T.}~\bibnamefont {Kottos}},
  \bibinfo {author} {\bibfnamefont {V.}~\bibnamefont {Kovanis}}, \ and\
  \bibinfo {author} {\bibfnamefont {D.~N.}\ \bibnamefont {Christodoulides}},\
  }\href {\doibase 10.1103/PhysRevA.85.013818} {\bibfield  {journal} {\bibinfo
  {journal} {Phys. Rev. A}\ }\textbf {\bibinfo {volume} {85}},\ \bibinfo
  {pages} {013818} (\bibinfo {year} {2012})}\BibitemShut {NoStop}%
\bibitem [{\citenamefont {Zhen}\ \emph {et~al.}(2015)\citenamefont {Zhen},
  \citenamefont {Hsu}, \citenamefont {Igarashi}, \citenamefont {Lu},
  \citenamefont {Kaminer}, \citenamefont {Pick}, \citenamefont {Chua},
  \citenamefont {Joannopoulos},\ and\ \citenamefont {Solja{\v c}i{\'
  c}}}]{zhen_spawning_2015}%
  \BibitemOpen
  \bibfield  {author} {\bibinfo {author} {\bibfnamefont {B.}~\bibnamefont
  {Zhen}}, \bibinfo {author} {\bibfnamefont {C.~W.}\ \bibnamefont {Hsu}},
  \bibinfo {author} {\bibfnamefont {Y.}~\bibnamefont {Igarashi}}, \bibinfo
  {author} {\bibfnamefont {L.}~\bibnamefont {Lu}}, \bibinfo {author}
  {\bibfnamefont {I.}~\bibnamefont {Kaminer}}, \bibinfo {author} {\bibfnamefont
  {A.}~\bibnamefont {Pick}}, \bibinfo {author} {\bibfnamefont {S.-L.}\
  \bibnamefont {Chua}}, \bibinfo {author} {\bibfnamefont {J.~D.}\ \bibnamefont
  {Joannopoulos}}, \ and\ \bibinfo {author} {\bibfnamefont {M.}~\bibnamefont
  {Solja{\v c}i{\' c}}},\ }\href {\doibase 10.1038/nature14889} {\bibfield
  {journal} {\bibinfo  {journal} {Nature}\ }\textbf {\bibinfo {volume} {525}},\
  \bibinfo {pages} {354} (\bibinfo {year} {2015})}\BibitemShut {NoStop}%
\bibitem [{\citenamefont {Cerjan}\ \emph {et~al.}(2016)\citenamefont {Cerjan},
  \citenamefont {Raman},\ and\ \citenamefont {Fan}}]{cerjan_zipping_2016}%
  \BibitemOpen
  \bibfield  {author} {\bibinfo {author} {\bibfnamefont {A.}~\bibnamefont
  {Cerjan}}, \bibinfo {author} {\bibfnamefont {A.}~\bibnamefont {Raman}}, \
  and\ \bibinfo {author} {\bibfnamefont {S.}~\bibnamefont {Fan}},\ }\href
  {\doibase 10.1103/PhysRevLett.116.203902} {\bibfield  {journal} {\bibinfo
  {journal} {Phys. Rev. Lett.}\ }\textbf {\bibinfo {volume} {116}},\ \bibinfo
  {pages} {203902} (\bibinfo {year} {2016})}\BibitemShut {NoStop}%
\bibitem [{\citenamefont {Chitsazi}\ \emph {et~al.}(2014)\citenamefont
  {Chitsazi}, \citenamefont {Factor}, \citenamefont {Schindler}, \citenamefont
  {Ramezani}, \citenamefont {Ellis},\ and\ \citenamefont
  {Kottos}}]{chitsazi_experimental_2014}%
  \BibitemOpen
  \bibfield  {author} {\bibinfo {author} {\bibfnamefont {M.}~\bibnamefont
  {Chitsazi}}, \bibinfo {author} {\bibfnamefont {S.}~\bibnamefont {Factor}},
  \bibinfo {author} {\bibfnamefont {J.}~\bibnamefont {Schindler}}, \bibinfo
  {author} {\bibfnamefont {H.}~\bibnamefont {Ramezani}}, \bibinfo {author}
  {\bibfnamefont {F.~M.}\ \bibnamefont {Ellis}}, \ and\ \bibinfo {author}
  {\bibfnamefont {T.}~\bibnamefont {Kottos}},\ }\href {\doibase
  10.1103/PhysRevA.89.043842} {\bibfield  {journal} {\bibinfo  {journal}
  {Physical Review A}\ }\textbf {\bibinfo {volume} {89}},\ \bibinfo {pages}
  {043842} (\bibinfo {year} {2014})}\BibitemShut {NoStop}%
\bibitem [{\citenamefont {Longhi}\ and\ \citenamefont
  {Valle}()}]{longhi_wrong_arxiv}%
  \BibitemOpen
  \bibfield  {author} {\bibinfo {author} {\bibfnamefont {S.}~\bibnamefont
  {Longhi}}\ and\ \bibinfo {author} {\bibfnamefont {G.~D.}\ \bibnamefont
  {Valle}},\ }\href@noop {} {\enquote {\bibinfo {title} {Loss-induced lasing:
  new findings in laser theory?}}\ }\bibinfo {note}
  {ArXiv:1505.03028}\BibitemShut {NoStop}%
\bibitem [{\citenamefont {Haus}(1983)}]{haus}%
  \BibitemOpen
  \bibfield  {author} {\bibinfo {author} {\bibfnamefont {H.~A.}\ \bibnamefont
  {Haus}},\ }\href@noop {} {\emph {\bibinfo {title} {Waves and {Fields} in
  {Optoelectronics}}}}\ (\bibinfo  {publisher} {Prentice Hall},\ \bibinfo
  {address} {Englewood Cliffs, NJ},\ \bibinfo {year} {1983})\BibitemShut
  {NoStop}%
\bibitem [{sup()}]{supp_mat}%
  \BibitemOpen
  \href@noop {} {}\bibinfo {note} {See supplementary material.}\BibitemShut
  {Stop}%
\bibitem [{\citenamefont {Figotin}\ and\ \citenamefont
  {Welters}(2012)}]{figotin_dissipative_2012}%
  \BibitemOpen
  \bibfield  {author} {\bibinfo {author} {\bibfnamefont {A.}~\bibnamefont
  {Figotin}}\ and\ \bibinfo {author} {\bibfnamefont {A.}~\bibnamefont
  {Welters}},\ }\href {\doibase 10.1063/1.4761819} {\bibfield  {journal}
  {\bibinfo  {journal} {J. Math. Phys.}\ }\textbf {\bibinfo {volume} {53}},\
  \bibinfo {pages} {123508} (\bibinfo {year} {2012})}\BibitemShut {NoStop}%
\bibitem [{\citenamefont {Figotin}\ and\ \citenamefont
  {Welters}(2014)}]{figotin_lagrangian_2014}%
  \BibitemOpen
  \bibfield  {author} {\bibinfo {author} {\bibfnamefont {A.}~\bibnamefont
  {Figotin}}\ and\ \bibinfo {author} {\bibfnamefont {A.}~\bibnamefont
  {Welters}},\ }\href {\doibase 10.1063/1.4884298} {\bibfield  {journal}
  {\bibinfo  {journal} {J. Math. Phys.}\ }\textbf {\bibinfo {volume} {55}},\
  \bibinfo {pages} {062902} (\bibinfo {year} {2014})}\BibitemShut {NoStop}%
\bibitem [{\citenamefont {T\"{u}reci}\ \emph {et~al.}(2006)\citenamefont
  {T\"{u}reci}, \citenamefont {Stone},\ and\ \citenamefont
  {Collier}}]{tureci06}%
  \BibitemOpen
  \bibfield  {author} {\bibinfo {author} {\bibfnamefont {H.~E.}\ \bibnamefont
  {T\"{u}reci}}, \bibinfo {author} {\bibfnamefont {A.~D.}\ \bibnamefont
  {Stone}}, \ and\ \bibinfo {author} {\bibfnamefont {B.}~\bibnamefont
  {Collier}},\ }\href@noop {} {\bibfield  {journal} {\bibinfo  {journal} {Phys.
  Rev. A}\ }\textbf {\bibinfo {volume} {74}},\ \bibinfo {pages} {043822}
  (\bibinfo {year} {2006})}\BibitemShut {NoStop}%
\bibitem [{\citenamefont {Ge}\ \emph {et~al.}(2010)\citenamefont {Ge},
  \citenamefont {Chong},\ and\ \citenamefont {Stone}}]{ge10}%
  \BibitemOpen
  \bibfield  {author} {\bibinfo {author} {\bibfnamefont {L.}~\bibnamefont
  {Ge}}, \bibinfo {author} {\bibfnamefont {Y.~D.}\ \bibnamefont {Chong}}, \
  and\ \bibinfo {author} {\bibfnamefont {A.~D.}\ \bibnamefont {Stone}},\
  }\href@noop {} {\bibfield  {journal} {\bibinfo  {journal} {Phys. Rev. A}\
  }\textbf {\bibinfo {volume} {82}},\ \bibinfo {pages} {063824} (\bibinfo
  {year} {2010})}\BibitemShut {NoStop}%
\bibitem [{\citenamefont {Cerjan}\ \emph {et~al.}(2015)\citenamefont {Cerjan},
  \citenamefont {Chong},\ and\ \citenamefont {Stone}}]{cerjan_csalt_2015}%
  \BibitemOpen
  \bibfield  {author} {\bibinfo {author} {\bibfnamefont {A.}~\bibnamefont
  {Cerjan}}, \bibinfo {author} {\bibfnamefont {Y.~D.}\ \bibnamefont {Chong}}, \
  and\ \bibinfo {author} {\bibfnamefont {A.~D.}\ \bibnamefont {Stone}},\
  }\href@noop {} {\bibfield  {journal} {\bibinfo  {journal} {Opt. Express}\
  }\textbf {\bibinfo {volume} {23}},\ \bibinfo {pages} {6455} (\bibinfo {year}
  {2015})}\BibitemShut {NoStop}%
\end{thebibliography}

%

\end{document}